\input harvmac
\newcount\figno
\figno=0
\def\fig#1#2#3{
\par\begingroup\parindent=0pt\leftskip=1cm\rightskip=1cm\parindent=0pt
\global\advance\figno by 1
\midinsert
\epsfxsize=#3
\centerline{\epsfbox{#2}}
\vskip 12pt
{\bf Fig. \the\figno:} #1\par
\endinsert\endgroup\par
}
\def\figlabel#1{\xdef#1{\the\figno}}
\def\encadremath#1{\vbox{\hrule\hbox{\vrule\kern8pt\vbox{\kern8pt
\hbox{$\displaystyle #1$}\kern8pt}
\kern8pt\vrule}\hrule}}

\overfullrule=0pt

%

\def\bar{\overline}

\def\T{{\bf T}}

\def\R{{\bf R}}

\font\zfont = cmss10 
\font\litfont = cmr6

\def\bigone{\hbox{1\kern -.23em {\rm l}}}
\def\ZZ{\hbox{\zfont Z\kern-.4emZ}}
\def\half{{\litfont {1 \over 2}}}

\def\hat{\widehat}

\Title{hep-th/9707093, IASSNS-HEP-97-83}
{\vbox{\centerline{ON THE CONFORMAL FIELD THEORY}
\bigskip
\centerline{OF THE HIGGS BRANCH }}}
\smallskip
\centerline{Edward Witten\foot{Research supported in part
by NSF  Grant  PHY-9513835.}}
\smallskip
\centerline{\it School of Natural Sciences, Institute for Advanced Study}
\centerline{\it Olden Lane, Princeton, NJ 08540, USA}\bigskip

\medskip

\noindent
We study $1+1$-dimensional theories of vector and hypermultiplets 
with $(4,4)$ supersymmetry.  Despite strong infrared fluctuations, these
theories flow in general to distinct conformal field theories on the Coulomb
and Higgs branches.  In some cases there may be  a quantum Higgs theory even
when there is no classical Higgs branch.
The Higgs branches of certain such theories provide
a framework for a matrix model of Type IIA fivebranes and the associated exotic
six-dimensional string theories.  Proposals concerning the interactions of
these string theories are evaluated.  
\Date{July, 1997}

\newsec{Introduction}

In the matrix model of $M$ theory \ref\bfss{T. Banks, W. Fischler, S. Shenker,
and L. Susskind, ``$M$ Theory As A Matrix Model: A Conjecture,'' 
 Phys. Rev. {\bf D55} (1997) 5112-5128,  hep-th/9610043 .},
there are a number of problems in which one must study supersymmetric
theories with eight supercharges that can arise by dimensional reduction
from six dimensions.  Such problems include $M$ theory with longitudinal 
fivebranes  \ref\bd{M. Berkooz and M. Douglas,  ``Fivebranes In M(atrix) 
Theory,''
Phys. Lett. {\bf B395} (1997) 196, hep-th/9610236 .}
 or in the presence of an ALE singularity 
\nref\md{G. Moore and M. Douglas,
``$D$-Branes, Quivers, and ALE Instantons,''
 hep-th/9603167.}\nref\douglas{M. Douglas,  ``Enhanced Gauge Symmetry
In M(atrix) Theory,'' 
 hep-th/9612126 .} \refs{\md,\douglas}.
To study these problems in uncompactified $M$ theory, one must consider the
dimensional reduction to $0+1$ dimensions of a six-dimensional theory
with vector and hypermultiplets.  
\nref\motl{L. Motl, ``Proposals On Nonperturbative Superstring Interactions,''
hep-th/9701025.}
\nref\bankseiberg{T. Banks and N. Seiberg, ``Strings From Matrices,''
hep-th/9702187.}
\nref\dvv{R. Dijkgraaf, E. Verlinde, and H. Verlinde, ``Matrix String
Theory,'' hep-th/9703030. }

Alternatively, instead of studying fivebranes or an ALE singularity in
 $M$ theory, one can compactify on a circle and consider
the same problems in weakly coupled Type IIA superstring theory.
To do so, one must generalize matrix string theory
\refs{\motl - \dvv}
of Type IIA superstrings to include hypermultiplets as
well as vector multiplets in the $1+1$-dimensional theory.

For closely related reasons,
the same sort of $1+1$-dimensional theory of vector and hypermultiplets 
governs 
the behavior of a D-onebrane parallel to a  D-fivebrane.  Investigations
in that context have shown \ref\dps{M. Douglas, J. Polchinski, and A. 
Strominger,
``Probing Five-Dimensional Black Holes With D-Branes,'' hep-th/97003031.}
that the interaction with the hypermultiplet induces on the Coulomb branch
(that is, on the vector multiplet moduli space) a metric with a rather peculiar
behavior -- a semi-infinite tube that appears at the ``origin'' in field space.
The metric with the semi-infinite tube is in fact the familiar, but
enigmatic, metric transverse to a fivebrane \nref\chs{C. G. Callan, Jr.,
J. A.  Harvey, and A. Strominger, ``Worldbrane Actions For String Solitons,''
 Nucl. Phys. {\bf B367} (1991) 60.}
\nref\rey{S.-J. Rey, ``On String Theory And Axionic Strings And Instantons,''
in {\it Particles And Fields '91} (Vancouver, 1991) 876.}
\refs{\chs,\rey}.  $1+1$-dimensional systems with several
vector and hypermultiplets give generalizations of the tube metric such as
an $A-D-E$ family \ref\seiberg{D. Diaconescu and 
N. Seiberg, ``The Coulomb Branch of (4,4) Supersymmetric Field Theories In
Two Dimensions,'' hep-th/9707158.}.

The Coulomb and Higgs branches 
of $1+1$-dimensional  theories of vector and hypermultiplets
are distinct quantum mechanically
\ref\witten{E. Witten, ``Some Comments On String Dynamics,'' hep-th/9507121,
{\it Strings '95} (World Scientific, 1996), ed. I. Bars. et. al.,
501.} despite strong $1+1$-dimensional infrared fluctuations.
The main goal of section 2 is to try to interpret
the tube metric on the Coulomb branch of vacua.  The tube metric 
requires some explanation
since the physics at the origin is surely non-singular.  The interpretation
that will be  proposed here is that the tube metric on the Coulomb 
branch is a manifestation
of a  conformal field theory to which the superrenormalizable theory
of vector and hypermultiplets can flow in the infrared.  This conformal
field theory is part of the Higgs branch if a classical Higgs branch exists;
otherwise it is a sort of quantum Higgs branch (at the ``origin'') that exists 
even
in the absence of a classical Higgs branch.

In section three we apply the results to matrix theory.  The main goal is to 
explore
the models relevant to ``matrix string theory'' of Type IIA fivebranes.   
First we
specialize to the particular cases of vector and hypermultiplets that are 
relevant
to the matrix theory of $k$ parallel longitudinal Type IIA fivebranes.
For the same reasons as in \refs{\motl - \dvv}, the limit in which the Type IIA
coupling vanishes is the infrared limit of the $1+1$-dimensional theory.
In this limit, the string coupling vanishes but interactions on the fivebrane
worldvolume survive \ref\newseiberg{N. Seiberg, ``New Theories In Six 
Dimensions,
And Matrix Description Of $M$ Theory On $T^5$ and $T^5/{\bf Z}_2$,''
hep-th/9705221.}  giving a fascinating
theory that is roughly a ``six-dimensional non-critical string theory.'' 
This six-dimensional theory 
corresponds to the Higgs branch of the $1+1$-dimensional theory.
We attempt to determine the extent to which
 a recent proposal \ref\newdvv{R. Dijkgraaf, E. Verlinde,
and H. Verlinde, ``5-D Black Holes And Matrix Strings,'' hep-th/9704018.}
can be used to describe the interactions in this six-dimensional theory; the 
results
are interesting but not entirely conclusive.

At an advanced stage of the present work, I learned of related work
\ref\ketal{O. Aharony, M. Berkooz, S. Kachru, N. Seiberg, and E. Silverstein,
``Matrix Description Of Interacting Theories In Six Dimensions,''
hep-th/9707079.} on the same models and especially
their  $0+1$-dimensional counterparts.  Related ideas were also developed
by M. Douglas, and I benefited from discussions with him.
 
\newsec{The Coulomb Branch And The Higgs Branch}
\subsec{Kinematics}

We begin with a six-dimensional supersymmetric theory with a gauge group $G$
and hypermultiplets  transforming in the representation $R$ of $G$.
We write $n_V$ and $n_H$ for the number of vector multiplets and 
hypermultiplets
(the dimension of $G$ and the quaternionic dimension of $R$, respectively)
and $r$ for the rank of $G$.  The  six-dimensional theory has an $SU(2)$ $R$ 
symmetry
$K$ which acts trivially on the gauge fields (and therefore non-trivially on 
their
fermionic partners) while the bosonic fields in the hypermultiplets transform
with $K=1/2$ (and their fermionic partners are invariant under $K$).

Upon dimensional reduction to $1+1$ dimensions, the six-dimensional
vector and hypermultiplets reduce to two-dimensional vector and
hypermultiplets.\foot{In two dimensions
there are $(4,4)$ multiplets other than the vector and hypermultiplets
\ref\gates{S. J. Gates, Jr. and S. V. Ketov, ``2D $(4,4)$ Hypermultiplets,''
hep-th/9504077.}, though they will not enter our discussion in the present
paper.}  The six-dimensional gauge field
splits into a two-dimensional gauge field plus scalars $\phi_i$, $i=1,\dots 
,4$ in
the adjoint representation of $G$.  The dimensional reduction produces an extra
$SO(4)$ $R$ symmetry group, under which the $\phi_i$ transform as a vector and
the scalars in the hypermultiplets are invariant.  We write this $SO(4)$ 
(or rather its double cover) as $L\times L'$, where 
$L$ and $L'$ are copies of $SU(2)$.

In the present section,
all Fayet-Iliopoulos $D$ terms and hypermultiplet bare masses are set to zero
and we assume that no hypermultiplets are neutral under $G$.   $n_H$ free 
hypermultiplets
would admit an action of a symmetry group $Sp(n_H)$, commuting with the $R$ 
symmetry
group $K$.
We write $H^{AX}$ for the scalars in the hypermultiplets; here $A=1,2$ is a 
$K$ index
(reflecting the fact that the $H$'s transform with $K=1/2$), and 
$X=1,\dots,2n_H$
is an $Sp(n_H)$ index.  
The gauge group $G$ is (via the representation $R$) a subgroup
of $Sp(n_H)$.  The $a^{th}$ generator of $G$, for $a=1,\dots, n_V$, is 
represented by
a tensor $t^a_{XY}$, and we write $\phi_i^a$, $a=1,\dots,n_V$ 
for the components of $\phi_i$.

At the classical level, a supersymmetric
state is simply a zero of the potential energy $V$ of the model.
$V$ is a sum of three terms which are multiples of
\eqn\junglo{\eqalign{ V_1& =\sum_{i<j}\Tr[\phi_i,\phi_j]^2   \cr
                      V_2& = \sum_{A,X,i}  |\sum_{a,Y}\phi_i^at^a_{XY}H^{AY}|^2
\cr
                      V_3& = \sum_{A,B,a}D^{ABa}D_{AB}^a.\cr}}
In the third line, $ D^{ABa}=t^a_{XY}H^{AX}H^{BY}$ 
is the hyper-Kahler moment map associated with the
action of $G$ on the $H$'s, preserving the hyper-Kahler structure. 

At a zero of $V$, all three terms separately vanish.
The vanishing of $V_1$ implies that the $\phi_i$ commute.
If the $\phi_i$ commute and are otherwise generic,
then vanishing of $V_2$ implies that $H=0$.
So there is always a ``Coulomb branch'' of classical vacua in which the 
$\phi_i$
are commuting and otherwise generic, and $H=0$.

If the representation $R$ is large enough, there is also a ``Higgs branch''
of vacua in which $H\not= 0$ and $\phi_i=0$.  There may also be mixed
Coulomb-Higgs branches in
which the $\phi_i$ are non-zero but not generic, and some hypermultiplets
are non-zero.  However, in this paper we consider only the
Coulomb and Higgs branches.

\subsec{ Separation Of The Two Branches}

In a similar situation above $1+1$ dimensions, the classical Coulomb and Higgs
branches just described are first approximations to quantum Coulomb
and Higgs branches, parametrizing families of quantum vacua.
In $1+1$ dimensions, because of strong infrared fluctuations of massless 
scalars,
one does not usually encounter continuous moduli spaces of quantum vacua.  
Roughly speaking, the ground state wave function on the Coulomb branch spreads 
over the whole Coulomb branch,  and likewise
the ground state wave function on the Higgs branch spreads over the whole
Higgs branch.  (In most instances  the branch in question is non-compact and 
there is 
actually
no normalizable quantum ground state wave function at all, but in any event the
quantum mechanical states spread over the Coulomb or Higgs branches.) 

Since the Coulomb and Higgs branches meet at $\phi=H=0$, one may question 
whether
the Coulomb and Higgs branches are actually
 distinct  quantum mechanically.  May not the wave function
spread from the Coulomb to the Higgs branch?  However, it has been argued
\witten\
that these branches really are distinct even in the quantum theory.
This follows upon considering the $(4,4)$ superconformal field theories 
that the $(4,4)$
supersymmetric gauge theory flows to in the infrared.  

A (4,4) superconformal field theory has an $SU(2)_L\times SU(2)_R$ group 
of left-moving
and right-moving $R$ symmetries.  Conformal symmetry implies that the $R$ 
symmetry
currents must be purely left-moving or purely right-moving, and hence cannot 
rotate
 scalar fields (a symmetry that rotates scalar fields
  cannot be separated locally into left- and right-moving pieces).\foot{
To state this argument very carefully, one should use the fact that generally
 the Coulomb
and Higgs branches are noncompact and consider regions at infinity that can
be analyzed semiclassically (all radii of curvature being large compared to
the string scale).   In such a region,  a symmetry that
rotates scalar fields cannot be locally separated into left- and right-moving
pieces.  In any model, therefore, in which the Coulomb or Higgs branches 
contain such
semiclassical regions, their $R$ symmetries cannot act on the massless
scalar fields of the branch in question.}
It follows then that the Higgs and Coulomb branches have separate 
$R$-symmetries.
The group $L\times L'$ described at the beginning of this section acts 
trivially
on the scalar fields $H^{AX}$ that are in the hypermultiplets, and so can be 
the
$R$ symmetry group of a (4,4) superconformal field theory obtained by infrared 
flow
on the Higgs branch.  Since $L$ and $L'$ act non-trivially on the scalars 
$\phi_i$ that
parametrize the Coulomb branch, they cannot appear as $R$ symmetries in a 
superconformal
field theory derived from the Coulomb branch.  The candidate $R$ symmetry 
of the Coulomb branch, 
because it acts
trivially on the $\phi_i$, is $K$ (this is only one $SU(2)$ symmetry; to 
get the expected
$SU(2)_R\times SU(2)_L$ $R$-symmetry group of a $(4,4)$ model, $K$ must 
break up in the infrared
as the sum of separately conserved left- and right-moving symmetries).  
As the Higgs and Coulomb branches have different $R$-symmetries, they must in 
fact be
different conformal field theories.

In most cases,
we could alternatively reason as follows.  The Coulomb branch is parametrized
by $r$ massless vector multiplets; all other fields are generically 
massive.  So the superconformal
field theory on the Coulomb branch has central charge
$\hat c=r$.\foot{In the usual examples of $(4,4)$ superconformal 
field theory, including
all examples that will be considered in the present paper, the 
left- and right-moving central
charges are equal; so we do not indicate a potential difference between 
them in the notation.  The
left- and right-moving Kac-Moody levels, introduced below, are likewise  
equal in these
models.}
On the other hand, if the representation $R$
is big enough so that the gauge group is generically completely broken 
on the Higgs branch, then
the Higgs branch is parametrized by $n_H-n_V$ massless hypermultiplets 
with all other fields
generically massive.  So the superconformal field theory on the Higgs 
branch has central charge 
\eqn\nomo{\hat c=n_H-n_V.}
In most examples, $r\not= n_H-n_V$, and this inequality of the central charges
shows that the Coulomb and Higgs branches are different.  (The example 
considered in
\witten\ was an exception with $r=n_H-n_V=1$.) 

Equation \nomo\ can be justified more generally by considering the $R$ 
symmetries
and {\it without} assuming that the gauge group is completely broken on the 
Higgs branch.
In $(4,4)$ superconformal field theory, $\hat c$ is equal to the ``level'' 
of the Kac-Moody
algebra of the left or right-moving $SU(2)$ group of $R$ symmetries.  This 
level can be computed
even away from criticality, for it equals the anomaly in the two   point 
function of the
$R$ symmetry currents.  The symmetry $L$, for example, couples to the 
left-moving
fermions in the hypermultiplets and the right-movers in the vector multiplets; 
so its
anomaly is $n_H-n_V$; and this is its Kac-Moody level in any possible 
conformal infrared limit
of the Higgs branch.  (Away from criticality, $L$ couples to left and 
right-moving bosons
in the vector multiplet; it becomes purely left-moving only in an infrared 
limit.)  
Hence a conformal field theory in which $L$ is an $R$-symmetry
has $\hat c = n_H-n_V$.

So we get a definition of a quantum Higgs branch that does not make any 
assumption about
breaking of the gauge symmetry at the classical level.  A (4,4) 
superconformal field theory
to which the gauge theory can flow in the infrared is a conformal 
field theory of the Higgs
branch if the $R$ symmetry group is $L\times L'$ and (therefore) 
the central charge is
$\hat c=n_H-n_V$.  Note that a quantum Higgs branch in this sense 
cannot exist if $n_H-n_V<0$.
If on the other hand $n_H-n_V=0$, then the quantum Higgs branch has $\hat c=0$ 
and
so is an infrared-trivial theory with a mass gap.  Since any theory 
with such a $\hat c=0$
quantum Higgs vacuum also has a quantum Coulomb branch with the same minimum
energy and no mass gap, 
such a quantum Higgs vacuum
with mass gap is a sort of ``bound state at threshold'' of the 
quantum field theory.  It is
somewhat analogous to bound states at threshold whose existence follows in 
certain
cases from string duality 
and can be proved via index theory \ref\sethistern{M. Sethi and M. Stern, 
``A Comment On The Spectrum Of $H$ Monopoles,'' Phys. Lett. {\bf B398} 
(1997) 47, hep-th/9607145, ``D-Brane Bound States Redux,''
hep-th/9705046.}.  It would be nice to know for sure if such quantum Higgs
vacua exist in two-dimensional (4,4) theories; decisive arguments one way
or the other will not be given in the present paper.

\subsec{Interpretation Of The Tube Metric}

Now we will focus on a specific case, which is $U(1)$ gauge theory 
with $k$ hypermultiplets.

At the classical level, the metric on the Coulomb branch of the $U(1)$ gauge 
theory
is simply the flat metric $|d\phi|^2/2e^2$.  A one-loop diagram with internal
hypermultiplets \dps\ induces on the Coulomb branch  a quantum metric 
\eqn\guffo{ds^2=|d\phi|^2\left({1\over 2e^2}+ {k\over 2|\phi|^2}\right).}
This metric is actually the same as the metric transverse to a Type IIA 
fivebrane
\refs{\chs,\rey}, a fact which is closely related to the fact that in matrix 
theory
one can interpret
\bd\ the inclusion of a hypermultiplet as the incorporation of a fivebrane in 
$M$ theory.

The metric in \guffo\ is enigmatic because there is a semi-infinite tube,
isometric to $\R\times {\bf S}^3$ with the obvious homogeneous metric, near
$\phi=0$.  In particular, $\phi=0$ is ``infinitely far away.''  This seems
strange because in the theory of vector and hypermultiplets, it does not seem
intuitively that $\phi=0$ is at an infinite distance.  Higher loop corrections 
to
\guffo\ cannot resolve the enigma, since \dps\ supersymmetry ensures that the
metric receives no further corrections.

It has been proposed \witten\ that the occurrence of such tube metrics is 
related
to the fact that the Coulomb and Higgs branches separate under renormalization
group flow.  
The intuitive idea was that $\phi=0$ (where the Coulomb and Higgs branches 
meet)
is ``infinitely far away'' by the time one flows to the infrared.
The precise form of the tube metric suggests, however, a further
and more specific interpretation.   The classical metric 
corresponds to a 
two-dimensional Lagrangian
\eqn\hoggo{ \int d^2x {|d\phi|^2\over 2e^2}}
that is conformally invariant if and only if $\phi$ has dimension zero, the 
usual conformal 
dimension for a scalar field in two dimensions.  But the tube metric is such 
that the 
Lagrangian near the origin
\eqn\noggo{k \int d^2x {|d\phi|^2\over 2|\phi|^2}} 
is conformally invariant for any conformal dimension that might be assigned 
to $\phi$.  
This strongly suggests that what is happening at the origin is that there 
is a conformal
field theory in which $\phi$ has a non-zero conformal dimension.
This is further suggested by  the fact \chs\ that $\ln|\phi|$ behaves as a 
Liouville
field in the theory in the tube.  

A simple inspection of \noggo\ does not reveal the conformal dimension of 
$\phi$,
but consideration of where \noggo\ came from makes it clear what the 
dimension must be.
\noggo\ was obtained by integrating out the hypermultiplets from an action that
is schematically
\eqn\imbo{\sum_{\alpha=1}^k\int d^2x \left(|(d+A)H_\alpha|^2+|\phi|^2|H_\alpha|
^2+
\bar\psi_\alpha\Gamma\cdot
(d+A)\psi_\alpha+\phi\bar\psi_\alpha\psi_\alpha + \lambda H_\alpha\psi_\alpha
 +D H_\alpha H_\alpha  \right).}
Here $H_\alpha$, $\psi_\alpha$, $\alpha=1,\dots,k$ are the bosons and fermions 
in 
the $\alpha^{th}$ hypermultiplet; 
the vector multiplet consists of the scalars $\phi$, gauge field
$A$, fermions $\lambda$, and auxiliary field $\vec D$, whose interactions
with the hypermultiplet are sketched in \imbo.
The full Lagrangian contains in addition to the hypermultiplet action \imbo\
the vector multiplet action which is schematically
\eqn\himbo{{1\over 2e^2}\int d^2x\left(|F|^2+|d\phi|^2+\bar \lambda 
\Gamma\cdot \partial \lambda +\vec D^2\right).}
$F=dA$ is the electric field.

Now \imbo\ is conformally invariant if and only if we assign dimension 
1 to $\phi$, $3/2$  to
$\lambda$, 1 to $A$, and 2 to $D$ and canonical dimensions (0 for $H$, 
$1/2$ for $\psi$) for
the hypermultiplets. 
These dimensions for the vector multiplet are not the usual canonical 
dimensions,
but they are ``geometrical'' dimensions; for instance, dimension one is the 
natural
geometrical dimension for a gauge field (this is clear from the fact that $A_i$
appears with $\partial/\partial x^i$ in  the covariant
derivative $D_i=\partial_i+A_i$), and the others are determined from this by 
supersymmetry.

 Since \noggo\ is obtained from \imbo\ by expanding
near constant $\phi$ and near zero momentum
and performing the path integral over the hypermultiplets,
conformal invariance of \noggo\ is a consequence of conformal invariance of 
\imbo.
If (and only if) we assign geometrical dimensions to the fields in the 
vector multiplet,
the entire hypermultiplet
path integral and not just the particular term extracted in \noggo\ 
will be conformally
invariant.

Once we assign dimensions $1,3/2,1,2$ to $\phi,\lambda,A,D$, the vector
multiplet kinetic energy \himbo\ is ``irrelevant'' (all terms have dimension 
four)
and can be dropped in flowing to the infrared.  Thus, we are dealing with
``induced gauge theory'' in which the vector multiplet kinetic energy 
comes completely
from performing the path integral over the hypermultiplets.  One might question
to begin with whether such an induced gauge theory actually makes sense.
At least for large $k$, one can argue as follows that it does.  The path 
integral
over the hypermultiplets induces a kinetic energy for the vector 
multiplets of the general
form
\eqn\hovvo{kL_{eff}(\phi,\lambda,A,D).} 
The overall factor of $k$ means that the effective theory of the vector 
multiplet
is weakly coupled for large $k$.  The expression $L_{eff}$ is conformally 
invariant
and has a non-degenerate expansion around $\phi=\lambda=A=D=0$, though -- 
as one would
expect in an interacting conformal field theory -- its expansion near zero 
momentum
is rather delicate.  The expansion in powers of $1/k$ can be made 
systematically using
propagators and vertices obtained by expanding $L_{eff} $ in powers of the 
fields.
No infinities arise in the perturbation expansion since there are no 
possible marginal
or relevant (dimension $\leq 2$) local gauge-invariant and 
supersymmetric counterterms in 
this theory.  (The expansion in powers of $1/k$ is somewhat analogous to the
$1/N$ expansion of the two-dimensional $(\bar\psi\psi)^2$ model above two 
dimensions,
which was  originally suggested in \ref\gn{D. J. Gross and A. Neveu,
``Dynamical Symmetry Breaking In Asymptotically Free Field Theories,''
Phys.Rev.D10:3235,1974.}.) 

The conformal field theory obtained in this way is, if $k$ is sufficiently 
large,
to be attributed to the Higgs branch and not the Coulomb branch. For in this
theory $H$ has dimension zero and can obtain an expectation value without 
breaking
conformal invariance, while $\phi$ has dimension one and cannot have an 
expectation value.
(On the Coulomb branch, of course, $\phi$ has dimension zero and can 
have an expectation
value.)  The central charge of this conformal field theory is $\hat c=k-1$ 
according
to \nomo.  The first term in the $1/k$ expansion would give the leading 
approximation
$\hat c\cong k$.  

Thus, for large $k$ the conformal field theory obtained in this way is not 
really
``new,'' but gives a useful way to describe the behavior of the Higgs
 branch near the origin.  Since the basic  puzzle of interpreting the 
tube metric of the Coulomb branch exists for all $k\geq 2$ \seiberg, 
it is natural to propose that this description
is valid all the way down to and including $k=2$.
It seems that the puzzle of the tube metric may not be present for $k=1$
\seiberg, and this suggests that for $k=1$ a quantum Higgs vacuum may not
be present.  Note that for $k=1$, a Higgs vacuum would have $\hat c=0$ and so
would correspond to a theory with mass gap that would be trivial
(and not just free) in the infrared limit.  Existence of a Higgs vacuum for
$k=1$ would be important in the theory of fivebranes, as will be clear in
section 3, so it would be desireable to obtain decisive arguments showing
that such a vacuum does or does not exist.

\subsec{Fixed Points in Superrenormalizable Gauge Theories}

We will try to put this discussion on somewhat firmer ground by observing
that somewhat similar phenomena occur in many superrenormalizable gauge 
theories
in two and three dimensions.

Consider a $U(1)$ gauge theory with gauge field $A$ and field strength $F$.  
Classically the object $I(\Sigma)=\int_\Sigma F$ is a topological
invariant, for any closed two-surface $\Sigma$ in spacetime.  One possible 
behavior of such
a theory is that the gauge field $A$ may decouple in the infrared limit.  If 
so, then
the different gauge bundles labeled by different values of $I(\Sigma)$ 
are indistinguishable
in the infrared and $F$ can have any dimension at all.  Usually what happens in
superrenormalizable theories is that $F$ has dimension less than 2 (for 
instance,
dimension 1 on the Coulomb branch of the supersymmetric two-dimensional
theories that we have been considering), in which case $I(\Sigma)$ 
diverges as one flows
to the infrared  and is not a well-defined  observable of the conformal field 
theory.

Alternatively, one may flow to an infrared fixed point at which $A$ does not 
decouple
from charged fields.  In that case, $F$ has dimension exactly 2, since this
is the only value compatible  with topological invariance of $I(\Sigma)$.

This discussion is probably not limited to abelian gauge theory.  
The intuitive idea is
that gauge fields of any gauge group either decouple in the infrared from 
charged
fields (including themselves in the non-abelian case) or have canonical 
dimensions. 
A more precise statement and argument is more difficult to give in the 
nonabelian
case because a gauge-invariant operator as convenient as $F$ does not exist.

We will illustrate these ideas with a number of examples.

\bigskip\noindent{\it QED In Two Dimensions}

First we consider two-dimensional QED, that is a $U(1)$ gauge theory in two
dimensions coupled to $k$ complex fermions $\psi_\alpha$, $\alpha=1,\dots ,k$
of the same charge.  The Lagrangian of this completely nonsupersymmetric model 
is
\eqn\hocco{L={1\over 4e^2}\int d^2x F_{ij}F^{ij}+ \sum_\alpha \int d^2x\,i 
\bar\psi_\alpha
\Gamma\cdot D\psi_\alpha.}
Imitating the discussion of the supersymmetric theory, we are led to suspect 
that at least for sufficiently large $k$, this theory
flows in the infrared to a conformal field theory that can be obtained by 
throwing
away the gauge kinetic energy and integrating out the fermions to induce an 
effective
kinetic energy.

This is in fact so for all $k$, as can be seen by bosonization.  One can 
replace
the fermions by bosons $\phi_\alpha$, $\alpha=1,\dots, k$, and the Lagrangian
by\foot{See \ref\coleman{S. Coleman, R. Jackiw, and L. Susskind,
``Charge Shielding And Quark Confinement In The Massive Schwinger Model,''
Ann. Phys. {\bf 93} (1975) 267;
S. Coleman, ``More On The Massive 
Schwinger Model,'' Ann. Phys. {\bf 101} (1976) 239.}
for this procedure and generalizations to analyze fermion mass terms.}
\eqn\ohocco{L'={1\over 4e^2}\int d^2x 
F_{ij}F^{ij} +\sum_{\alpha=1}^k\int d^2x
\left(\half |d\phi_\alpha|^2+{\phi_\alpha\epsilon^{ij}F_{ij}\over \sqrt 
\pi}\right).}
This flows as expected to a conformal field theory in which $F$ has dimension 
2,
$\phi$ has dimension 0, and the original $F^2$ kinetic energy is irrelevant in 
the
infrared and can simply be dropped.  

The central charge can be computed using the fact that one combination of the 
scalars
(namely their sum) combines with $F$ to a massive field (the mass is infinite
if one discards the $F^2$ term from the Lagrangian), while the other $k-1$
scalars are free and 
massless.  So in fact $c=k-1$.  At $k=1$, since $c=0$, we have a theory
with a mass gap.  In fact, this theory is the Schwinger model, long known to be
equivalent by the above procedure to the theory of a massive free boson.  The 
massive
theory at $k=1$ is analogous to the massive Higgs vacuum that might possibly
exist for the supersymmetric
$U(1)$ theory with one hypermultiplet, except that (as there is no 
Coulomb branch) it is not embedded in a continuum.  

One oddity of the present model is that, while the correlation 
functions are conformally
invariant in the infrared if $F$ is assigned dimension two, connected 
correlation
functions involving $F$ have no infrared singularities.  This would not be so 
in the supersymmetric example considered earlier if $k>1$; it is 
true, of course, for $k=1$
as there is a mass gap and no correlation functions have infrared 
singularities.

The $1/k$ expansion has a similar qualitative structure in the 
nonabelian case.  For this,
one replaces the gauge group by a nonabelian group and takes the fermions to 
consist
of $k$ copies of any given representation.  The Lagrangian is 
formally just like \hocco\
(except that of course the gauge kinetic energy is now $\Tr F^2$) 
and there is a large
$k$ expansion as in the abelian case.  For
a recent brief discussion see \ref\sonn{J. Sonnenschein, 
``More On Screening And Confinement
In 2D QCD,'' hep-th/9707031.}. 
\nref\gaw{K. Gawedzki and A. Kupianen, ``A $G/H$ Conformal Field Theory From
Gauged WZW Model,'' Phys. Lett. {\bf 215B} (1988) 119, ``Coset Construction
From Functional Integrals,'' Nucl. Phys. {\bf B320 (FS)} (1989) 649.}
\nref\mart{G. Guadagnini, M. Martellini, and M. Mintchev, ``Scale-Invariant
Sigma Models On Homogeneous Spaces,'' Phys. Lett. {\bf 194B} (1987) 69.}
\nref\bark{K. Bardakci, E. Rabinovici, and B. Saring, 
``String Models With $c<1$ Components,''
Nucl. Phys. {\bf B299} 
(1988) 157; A. Altschuler, K. Bardakci, and E. Rabinovici, ``A Construction
Of The $c<1$ Modular Invariant Partition Functions,''
 Comm. Math. Phys.
{\bf 118} (1988) 157;  E. Rabinovici, ``Aspects Of A Lagrangian Formulation
For Modular-Invariant Coset Constructions,'' in {\it Conformal Field Theories
And Related Topics}, proceedings, Annecy 1988, 192.}
\nref\schn{D. Karabali, Q.-H Park, H. J. Schnitzer, and Z. Yang,
``A GKO Construction Based On A Path Integral Formulation Of Gauged 
Wess-Zumino-Witten Actions,''
Phys. Lett. {\bf 216B} (1989) 307; H. J. Schnitzer, ``A Path Integral 
Construction
Of Superconformal Field Theories From A Gauged Supersymmetric 
Wess-Zumino-Witten
Action,''
Nucl. Phys. {\bf B324}
(1989) 412; D. Karabali and H.-J. Schnitzer, ``BRST Quantization Of The Gauged
WZW Action And Coset Conformal Field Theories,''
Nucl. Phys. {\bf B329} (1990) 649.}
 One can even extend the discussion to small $k$
 via the relation between gauged WZW models and coset conformal
field theories \refs{\gaw-\schn}.   To do so, let $k'$ be the total number 
of real fermi fields
in the model, summed over representations.  One 
can replace the fermions by an $SO(k')$ WZW model 
at level one.
Upon deleting the gauge kinetic energy from the Lagrangian, one 
gets a gauged WZW model
which flows in the infrared to a coset conformal field theory.  Using the 
machinery
of coset conformal field theory, one can argue that the gauge 
fields have their canonical
dimension in this infrared fixed point and one can describe the 
fixed point quantitatively.

\bigskip\noindent{\it QED In Three Dimensions}

\nref\ar{T. Appelquist and R. Pisarski, Phys. Rev. {\bf D23} (1981) 2305.}
\nref\aph{T. Appelquist and U. Heinz, ``Three-Dimensional $O(N)$  Theories At 
Large
Distances,'' Phys. Rev. {\bf D34} (1981) 2169.}
\nref\rp{R. Pisarski, Phys. Rev. {\bf D29} (1984) 2423.}
\nref\morea{T. Appelquist, M. Bowick, E. Cohler, and L. C. R. Wijewadhana,
Phys. Rev. Lett. {\bf 55} (1985) 1715, T. Appelquist, M. Bowick, D.
Karabali, and L. C. R. Wijewardhana, Phys. Rev. {\bf D33} (1986) 3704.}
\nref\omorea{T. Appelquist, D. Nash, and L. C. R. Wijewardhana, ``Critical 
Behavior
In $(2+1)$-Dimensional QED,'' Phys. Rev. Lett. {\bf 60} (1988) 2575.}
\nref\an{T. Appelquist and D. Nash, ``Critical Behavior In $(2+1)$-Dimensional
QED,'' Phys. Rev. Lett. {\bf 60} (1990) 721.}
Next we consider the same  model -- $U(1)$ coupled to $k$ species of complex
fermion -- but now in three dimensions.  The $1/k$ expansion of this theory 
has been
extensively studied over a period of many years, some of the references being
\refs{\ar-\an}.  In particular, it is demonstrated that order by order in 
$1/k$,
there is a non-trivial infrared fixed point at which $F$ has dimension two and 
the
ordinary kinetic energy is irrelevant.  
The fact that the dimension of $F$ is precisely two to all orders in $1/k$ is 
argued
in \an.  A difference from the two-dimensional case is that it has been argued
\refs{\omorea,\an} that the flow to a non-trivial infrared fixed point of this 
nature
occurs only for $k$ greater than a certain critical value.

As in two dimensions, one can generalize to the nonabelian case without 
changing
the qualitative structure of the $1/k$ expansion.  In fact, the 
argument in \an\ showing
dimension 2 for $F$ is made in this context.

\bigskip\noindent{\it Supersymmetric QED In Three Dimensions}

For our last such example, we consider supersymmetric QED in four dimensions.
For brevity we consider only the case of $N=4$ supersymmetry 
(that is, eight supercharges)
though one could likewise consider $N=2$ supersymmetry 
\ref\iasru{
O. Aharony, A. Hanany, K. Intriligator, N. Seiberg, and M. J. Strassler,
``Aspects Of $N=2$ Supersymmetric Gauge Theories In Three-Dimensions,''
hep-th/9703110.}
to give further examples.

\nref\seidthree{N. Seiberg, ``IR Dynamics Of Branes And Compactification
To Three Dimensions,'' Phys. Lett. {\bf B384} (1996) 81, hep-th/9606017.}
\nref\swdthree{N. Seiberg and E. Witten, ``Gauge Dynamics And Compactification
To Three Dimensions,'' in {\it The Mathematical Beauty Of Physics},
ed. J. M. Drouffe and J. B. Zuber (World-Scientific, 1997) 333,
 hep-th/9607163.}
So we consider a three-dimensional $U(1)$ gauge theory of a vector multiplet 
coupled to $k$ hypermultiplets.  On the Coulomb branch there is a gauge field
$A$ and three scalars $\vec\phi$.  
After dualizing the gauge field to a fourth scalar,
the Coulomb branch is described by a Taub-NUT hyper-Kahler
metric with an $A_{k-1}$ orbifold singularity \refs{\seidthree,\swdthree}.
The Taub-NUT metric corresponds 
 in terms of the original variables $\vec\phi$, $A$, to a  kinetic energy 
which looks like
\eqn\polly{k\int d^3x\left({|d\vec\phi|^2+F^2\over |\vec\phi|}\right)}
(with just such coefficients as to produce an $A_{k-1}$ orbifold singularity
after duality).  We see the familiar fact that conformal invariance holds if 
and only
if the fields are assigned their geometrical dimensions, namely 2 for $F$ and 
$1$ for $\vec\phi$.
These are the dimensions that make the hypermultiplet kinetic energy 
conformally
invariant, and if one did not have exact methods to analyze this problem, one 
could
arrive at \polly\ as the behavior near the origin for large $k$
by simply integrating out the hypermultiplets.
Since there are exact methods, one can be precise about how large
$k$ must be to produce such behavior at the origin.  The form \polly\ for the
metric near $\vec\phi=0$ is valid for all $k\geq 1$.  For $k\geq 2$, this
behavior at the origin actually corresponds to a critical point with 
non-trivial
infrared behavior, while for $k=1$, a duality transformation (to the scalar
field dual to $A$) eliminates the singularity, giving a smooth Taub-NUT metric.
So in that case, the conformal field theory at the origin is actually
infrared-free (but non-trivial, that is there is no mass gap) 
if expressed in the right variables.

\bigskip\noindent{\it QED In Four Dimensions}

We conclude by considering a model that is {\it not} superrenormalizable, 
namely
QED in four dimensions.  In this case, the fact that $F$ must have dimension 
two
at a critical point at which the photon does not decouple from electrons
implies \ref\lots{
P. C. Argyres, M. R. Plesser, N. Seiberg, and E. Witten,
``New $N=2$ Superconformal Field Theories In Four Dimensions,''
Nucl. Phys. {\bf B461} (1996) 71, hep-th/9511154.}
that $U(1)$ gauge theory
in four dimensions cannot have such an infrared fixed point.
This argument holds only in the absence of magnetic monopoles, since the 
Bianchi identity, which is used to show
that $I(\Sigma)$ is topologically invariant and
$F$ must have dimension 2, holds only in the absence of monopoles.

\newsec{Matrix Theory With Hypermultiplets}

\subsec{Models}

In this section, we consider matrix string theory with hypermultiplets.

To incorporate longitudinal fivebranes we follow \bd.  Type IIA matrix string 
theory
is based on a $U(n)$ gauge theory with $(8,8)$ supersymmetry.  From the point 
of view
of $(4,4)$ supersymmetry, this model has a $U(n)$ vector multiplet and a 
hypermultiplet,
which we will call $X$, in the adjoint representation of $U(n)$.  To 
incorporate
$k$ parallel longitudinal fivebranes, one must add  $k$ hypermultiplets
$H_\alpha$, $\alpha=1,\dots,k$ in the fundamental representation of $U(n)$.
\foot{What is considered in \bd\ is the problem of adding a hypermultiplet
 to zerobrane
quantum mechanics to represent an $M$ theory fivebrane.  To make a Type IIA
 construction,
one considers, in a standard fashion, a periodic array
of fivebranes. After making a sort of Fourier transform or $T$-duality 
transformation
\ref\taylor{W. Taylor IV, ``D-Brane Field Theory On Compact Spaces,''
Phys. Lett. {\bf B394} (1997) 283, hep-th/9611042.},
one arrives at the $1+1$-dimensional field theory with hypermultiplets 
representing
fivebranes.}  Lorentz-invariant physics hopefully emerges in the large $n$ 
limit.

Alternatively,
to describe matrix string theory in the presence of an $A-D-E$ singularity,
one considers \refs{\md,\douglas} a certain collection of unitary groups and 
hypermultiplets
associated with the $A-D-E$ Dynkin diagram.  

In the $A-D-E$ case, the Fayet-Iliopoulos or FI
couplings are quite important and lead
to resolution of the   singularity.   In the presence of generic FI terms, the 
model
has a Higgs branch and no Coulomb branch.  The Higgs branch is interpreted 
physically
in terms of motion of supergravitons in spacetime.  
The physical interpretation of the Coulomb branch 
that occurs
when the FI terms vanish is an intriguing question that will not be considered 
here.
We will focus instead on fivebranes.

\nref\ustrom{A. Strominger, ``Open $p$-Branes,'' Phys. Lett. {\bf B383} (1996) 
44,
hep-th/9512059.} 
In studying fivebranes, the goal is to understand the intrinsic fivebrane 
theory
obtained \newseiberg\ by taking the Type IIA string coupling constant to zero, 
whereupon
a residual six-dimensional theory survives.  Upon a further reduction 
($\alpha'\to 0$)
this theory reduces to the exotic six-dimensional field theory found earlier 
\refs{\witten,\ustrom}.
As in conventional Type IIA matrix string theory \refs{\motl - \dvv}, 
the weak coupling limit is obtained by formulating the $1+1$-dimensional 
theory on
a circle ${ S}$ of radius $R$, and taking $R\to\infty$.  The large $ R $
behavior is controlled by the possible conformal
field theories to which the $1+1$-dimensional gauge theory can flow.

An important point is that an intrinsic interacting fivebrane theory in
the limit that the  Type IIA string coupling constant goes to zero is only
known to exist for the case of $k\geq 2$ parallel fivebranes.  The limiting
theory must be non-trivial in this case since even in the infrared limit
it flows to the non-trivial theory described in \refs{\witten,\ustrom}.
In the analogous Type IIB case the argument that the fivebrane theory remains
interacting as the string coupling constant goes to zero \newseiberg\
uses the fact that
the gauge theory on the fivebrane world-volume has a $U(k)$ gauge group, which
is non-abelian for $k\geq 2$.  Thus, one only has a firm argument that an
intrinsic fivebrane theory exists in the case $k\geq 2$.  Because this theory
seems to be the $A_{k-1}$ case of an $A-D-E$ story, and $A_{k-1}$ corresponds
to $SU(k)$ rather than $U(k)$, one might suspect that the $U(1)$ is completely
decoupled (and not just decoupled in the infrared limit) and that there is no
intrinsic theory in this sense for a single fivebrane.

The $U(n)$ theory with an adjoint and $k$ fundamental hypermultiplets 
$H_\alpha$,
$\alpha=1,\dots,k$, has
both a Coulomb branch and a Higgs branch.  (The classical Higgs branch has 
rather special
properties for $k=1$, as we discuss later, and is generic for $k>1$.)  As 
is clear from the
metric \guffo, the motion on the Coulomb branch describes the motion 
perpendicular
to the fivebrane.  In fact, near infinity on the Coulomb branch, the 
$H_\alpha$
can be neglected (their masses become large), and one reduces to Type IIA 
matrix
string theory in the absence of fivebranes.

So the fivebranes must be described by the conformal field theory of the Higgs
branch.  On this branch, the scalars in the $U(n)$ vector multiplet are near
zero, and the transverse
space of the light cone matrix string theory is reduced from eight to 
four dimensions,
the correct number to describe  a six-dimensional theory in light cone gauge.

Whenever there exists an intrinsic fivebrane theory that survives for vanishing
string coupling,  the conformal field theory
of the Coulomb branch must have some pathology.    For
if the Coulomb branch is governed by a completely well-behaved $(4,4)$ 
superconformal field theory (with no problems except the standard
noncompactness at infinity),
then this conformal field theory is the starting point for a systematic
Type IIA perturbation expansion.  It would surely be inconsistent to add
fivebranes by hand to such a perturbation expansion.  So (a) if the Coulomb
branch conformal field theory is completely well-defined, it must give
a completely self-contained description of physics in the field of the
fivebrane and hence the usual modes
propagating on the fivebrane world-volume must be described by
ordinary vertex operators
in this theory  (as was assumed in the earliest work on fivebrane conformal
field theory \chs), and (b) if there is a limiting non-trivial fivebrane theory
in the limit of zero string coupling constant (as we expect at least for
$k\geq 2$) the Coulomb branch conformal field theory cannot be completely
well-defined.

We independently expect a problem with the Coulomb branch for $k\geq 2$
because of the tube metric (which apparently exists for $k\geq 2$ \seiberg)
whose existence leads to a failure of normalizability of the Coulomb branch
vacuum near the fivebrane location (near $\phi=0$ in the language of section 
two)
and a blow-up of the effective Type IIA string coupling constant.    
The problem with the Coulomb branch when there is an intrinsic fivebrane theory
is analogous to the fact that one cannot expect to give a completely consistent
description of the physics outside a black hole without including
black hole degrees of freedom.  

For $k>1$, after specifying the orientation of an $M$ theory fivebrane, five 
additional real numbers are needed to specify its transverse position.  In 
\bd, 
these parameters were interpreted as bare masses of the $H_\alpha$, or more 
exactly 
as relative bare masses.
To be more precise, in the $0+1$-dimensional
construction in \bd, the $\alpha^{th}$ hypermultiplet interacts not with the 
scalars
$\vec\phi$ in the vector multiplet but with $\vec\phi+\vec a_\alpha$,    where
(as $\vec \phi$ has five components in $0+1$ dimensions) each $\vec a_\alpha$ 
is a five-component
``transverse position vector'' for the $\alpha^{th}$ fivebrane. 
Addition of an overall constant
to all $\vec a_\alpha$ can be absorbed in adding a constant to $\vec \phi$ 
(just as
an overall constant in the positions of a collection of fivebranes can be 
absorbed in a 
translation).  

In $1+1$-dimensions, the $\vec\phi$ and $\vec a_\alpha$ have only four 
components;
the fifth component of the fivebrane position has a different 
interpretation that we discuss
momentarily.  The hypermultiplet bare masses $\vec a_\alpha$
are relevant operators, of dimension one.
If therefore we want these perturbations to survive and not dominate as $R$ 
(the radius of
the circle ${ S}$) becomes large, we must
take the $\vec a_\alpha$ to be of order $1/R$ as $R\to\infty$.  

What about the fifth component of the fivebrane position?  After the 
$T$-duality to
a $1+1$-dimensional model, this component becomes a Wilson line around ${ S}$.
The $U(n)$ gauge field with which $H_\alpha$ interacts can be supplemented by 
an
$\alpha$-dependent
additive constant, which is such that the holonomy around $S$ contains 
an $\alpha$-dependent
multiplicative factor $e^{i\theta_\alpha}$, where $\theta_\alpha$ is the 
position
of the fivebrane in the eleventh dimension -- the dimension that was 
compactified to go
from $M$ theory to Type IIA superstrings.  The $\theta_\alpha$ should be held
fixed as $R\to\infty$.

\subsec{ Analysis Of The Classical Higgs Branch}

We will now analyze the classical structure of the Higgs branch 
${\cal M}_{n,k}$ of the $U(n)$ theory with an adjoint and $k$ fundamental
hypermultiplets.
The results will be useful later when we consider interactions.

The Higgs branch of this theory
(with zero bare masses and FI terms) can be interpreted as the moduli space
of $n$-instanton solutions of $U(k)$ gauge theory on 
${\bf R}^4$,
 partially compactified by incorporating
small instantons.\foot{In this paper, whenever we speak of instanton moduli 
spaces on
${\bf R}^4$, we always have in mind based instantons; that is, in defining
the moduli space one considers two instantons to be gauge-equivalent if and
only if they
are equivalent by a gauge transformation that equals the identity at 
infinity.}  
This fact follows from the ADHM construction of 
instantons, and is 
important in many of the physical interpretations of the model.
For $k=1$, as $U(1)$ is abelian, there are no ``honest'' instantons,
and all solutions correspond to completely collapsed instantons.  
The completely collapsed
instantons correspond to $H=0$, so are represented by the values 
of the adjoint hypermultiplet
only.   We will momentarily give a direct proof, without using the
relation to instantons, that $H=0$ in any vacuum
of the $k=1$ theory.

To analyze the generic structure of the Higgs branch for any $k$,
we follow the concluding portion of
\ref\hurtu{J. Hurtubise, ``Instantons And Jumping Lines,''
Comm. Math. Phys. {\bf 105} (1986) 107.} and proceed as follows.
Pick a complex structure on the four-dimensions of the light-cone fivebrane 
world-volume and view
the $(4,4)$ model
 as a $(2,2)$ supersymmetric model.  Under this reduced supersymmetry,
the adjoint hypermultiplet $X$ splits
into a pair $U,V$ of chiral multiplets in the adjoint representation.  
The $H_\alpha$, $\alpha=1
,\dots,k$,
become chiral multiplets $A_\alpha$, $B_\alpha$ in the fundamental 
and antifundamental
representations of $U(n)$.  The vanishing of the $(4,4)$ 
$\vec D$     terms give a complex equation
\eqn\olpo{[U,V]^i{}_j+\sum_\alpha A^i_\alpha B_{\alpha\,j}=0}
together with a real equation which is the vanishing of the $(2,2)$ $D$ field, 
which
we call $D_{\bf R}$.  The Higgs branch is described by solving 
\olpo, setting $D_{\bf R}$ to
zero, and dividing by $U(n)$.  If things are sufficiently generic 
(which in the present problem
is true for $k>1$ but not for $k=1$), then setting $D_{\bf R}$ to zero and  
dividing
by $U(n)$ is equivalent on a dense open set to dividing by the complexification
$GL(n,{\bf C})$ of $U(n)$.  

Now we proceed to solve \olpo\ and divide by $GL(n,{\bf C})$.  
Generically we can diagonalize $U$, say with eigenvalues $u_i$, $i=1,\dots 
,n$.  
Diagonalizing $U$ fixes most of the $GL(n,{\bf C})$ symmetry; what remains are
the diagonal $GL(n,{\bf C})$ transformations -- forming a group 
$({\bf C}^*)^n$ -- and
the Weyl group -- which acts by permutation of the eigenvalues.
Equations \olpo\ put no restrictions on the diagonal matrix elements 
$v_i=V^i{}_i$.
The Higgs branch will thus be parametrized among other things 
by the values of the $u_i$
and the $v_i$.  One can generically solve \olpo\ for the off-diagonal 
$V^i{}_j$ via
\eqn\holpo{V^i{}_j=-{A^iB_j\over u_i-u_j},~~{\rm for}~i\not= j.}
Finally, vanishing of the diagonal terms in \olpo\ puts a 
restriction on  $A$ and 
$B$, which is that
\eqn\mumbo{\sum_\alpha A^i_\alpha B_{\alpha\,i}=0,\,\,{\rm for}\,i=1,\dots,n,}
with  no sum over $i$ in this equation.

The surviving data, for each fixed $i$, are the diagonal matrix elements
$u_i$ and $v_i$, together with the hypermultiplets 
$A^i_\alpha$, $B_{\alpha\,i}$.
$A^i_\alpha$ and $B_{\alpha\,i}$ must obey \mumbo\ and one must impose an 
equivalence
relation $A\to \lambda A$, $B\to \lambda^{-1}B$ that comes from the 
residual ${\bf C}^*$
symmetry.  The data $u,v,A,B$ with this condition and this equivalence relation
make up a copy of the Higgs branch ${\cal M}_{1,k}$ for $n=1$ with $k$ 
hypermultiplets.
(By the ADHM construction, this is the same as the one-instanton moduli
 space for the group
$U(k)$.) Since we get $n$ copies of this space, one for each $i$, and we then
must divide by the Weyl group, which acts by permutations of the eigenvalues, 
the conclusion
is that on a dense open set the desired Higgs moduli space ${\cal M}_{n,k}$ 
is the symmetric
product $S^n{\cal M}_{1,k}$ of $n$ copies of ${\cal M}_{1,k}$.  Technically,
${\cal M}_{n,k}$ is {\it birational} to $S^n{\cal M}_{1,k}$.  Note that this
identification of ${\cal M}_{n,k}$ with $S^n{\cal M}_{1,k}$ does not respect 
all of the
symmetries of the problem; in fact, it depended on the choice at the outset 
of a complex structure.

Now we look more closely at the special case $k=1$.  In this case, there 
is no sum over
$\alpha$ in \mumbo, which collapses to the condition that $A^iB_i=0$ for each 
$i$.
So after a rearrangement of the eigenvalues, we can assume that $A^i\not= 0$, 
$B_i=0$,
for $i=1,\dots,n_1$, $A^i=0,B_i\not= 0$ for $n_1+i\leq i\leq n_2$, and 
$A^i=B_i=0$
for $i>n_2$.  In this basis, \holpo\ shows that $V^i{}_j$ 
is upper triangular.  An
examination of the equation $D_{\bf R}=0$ now shows that it can be satisfied 
only for
$A^i=B_i=0$.
(Thus this is an exceptional case in which setting $D_{\bf R}=0$ 
and dividing
by $U(n)$ is not generically equivalent to dividing by $GL(n,{\bf C})$, because
``the generic orbit is not stable.'')  This confirms the claim made at the 
outset of the
present discussion that for $k=1$, the  hypermultiplet $H$
is identically zero on the Higgs branch.  \olpo\ now reuires that $U$ 
and $V$ commute,
so the Higgs branch is described by the eigenvalues $u_i$, $v_i$.  (The 
vanishing of
$D_{\bf R}$ forbids solutions in which $U$ and $V$ have Jordan forms looking 
like
$\left(\matrix{ \lambda & 1\cr 0 & \lambda}\right)$ and cannot be 
diagonalized.)
In particular, the moduli space ${\cal M}_{1,1}$ is parametrized by a single 
pair
of complex numbers $u,v$, and is a copy of ${\bf R}^4$.  More 
generally, 
${\cal M}_{n,1}$ is parametrized by $n$ pairs $u_i,v_i$ up to permutation, and 
so
is isomorphic to a symmetric product 
\eqn\murr{{\cal M}_{n,1}=S^n{\cal M}_{1,1}.}

There are a few differences between \murr\ and the corresponding and 
superficially
similar relation between ${\cal M}_{n,k}$ and ${\cal M}_{1,k}$:

(1) The relation \murr\ was deduced from the exact statement that $H=0$ in any
vacuum for $k=1$ and is true on the nose as a relation between the spaces 
involved;
no blowup or birational transformation is involved.
The fact that for general $n$
the birational  equivalence of ${\cal M}_{n,k}$ with $S^n{\cal M}_{1,k}$ 
involves
the equation \holpo\ which has a pole at $u_i=u_j$ shows that in the latter 
case we are only getting a birational transformation, which breaks down
when $u_i=u_j$.

(2) Related to this, the hyper-Kahler metric on the moduli space ${\cal 
M}_{n,1}$ is
the obvious flat hyper-Kahler metric on the symmetric product $S^n{\bf R}^4$.  
This
is deduced by setting $A=B=0$ and evaluating the classical Lagrangian for 
diagonal (but spacetime-dependent) matrices $U,V$.  By contrast, since \holpo\ 
shows
that the off-diagonal matrix elements of $V$ do {\it not} vanish for $k>1$,
the metric on ${\cal M}_{n,k}$ does not have such an elementary relation to 
that
on ${\cal M}_{1,k}$ for $k>1$.

(3) Finally, \murr\ is a consequence of the fact that $H=0$ on the 
Higgs branch of the
$k=1$ theory, which is a statement completely invariant under all symmetries 
of 
the problem,
though our proof of it was not invariant.  So the identification \murr\ is 
completely
invariant.  That is not so for the corresponding birational equivalence for $k>
1$, 
which depended on a choice of complex structure.

\subsec{The Hamiltonian}

Now we will try to make as explicit as possible the structure of the
theory, first in the case $k=1$.

We have seen in the last subsection that the Higgs branch for $k=1$ 
is parametrized by the adjoint hypermultiplet $X$, with $H=0$.  $X$ contains
 four scalar fields $X_\lambda, \,\lambda=1,\dots,4$ in the adjoint 
representation
of $U(n)$.  A point on the Higgs branch is parametrized by  $X$'s that
mutually commute, modulo the action of $U(n)$.  The simultaneous eigenvalues of
the $X_\lambda$ define a set of $n$ points in ${\bf R}^4$, which are uniquely
determined up to the action of the Weyl group, that is, up to permutation.

If one is far out on the Higgs branch, that is if the $n$ points in ${\bf R}^4$
are all far apart, then the gauge group $U(n)$ is spontaneously broken to
$U(1)^n$.  At this point, the low energy theory looks like $n$ copies of 
the $n=k=1$ system.  This system consists of $U(1)$ with a single charged
hypermultiplet plus an adjoint-valued hypermultiplet, which, as $U(1)$ is 
abelian,
is simply free.  It is possible that the $U(1)$ theory with one
charged hypermultiplet has no quantum Higgs vacuum
and has a completely well-defined Coulomb branch conformal field theory.
In this case, upon including also the free hypermultiplet from the adjoint,
we get a theory which has only one quantum branch in which the scalars
in the vector multiplet and the free hypermultiplet are both free to vary.
This theory would describe both the physics outside the fivebrane and the
fivebrane modes, so there would be no separate fivebrane theory that can be
decoupled from the theory in bulk.  This would have to be a completely 
well-defined
conformal field theory (except for the usual subtleties coming from the 
noncompactness at infinity), and the $k=1$ fivebrane modes would be described
by ordinary vertex operators in this theory.

On the other hand, it is conceivable
that $U(1)$ with one charged hypermultiplet can flow
in the infrared to a massive Higgs vacuum.  If so, then the $n=k=1$ system
has a quantum Higgs branch in which only the scalars in the free hypermultiplet
can vary.  This branch will lead to the existence of an
intrinsic theory for a single Type IIA fivebrane.

We will proceed with the discussion of interactions assuming that such
a branch exists for $n=k=1$.  Even if it turns out that this is not so,
the discussion of this case will give useful practice for $k>1$ where there
definitely is an intrinsic fivebrane theory.

So now we consider the case $k=1$, $n>>1$.  Far out on the Higgs branch,
the eigenvalues of the scalars in the adjoint hypermultiplet define
$n$ points in ${\bf R}^4$ (and the scalars in the vector multiplet are locked
near zero because of our assumptions about the $n=k=1$ system). 
The key question is what happens when  some of the $n$ points in ${\bf R}^4$
come near by.  We have determined that the Higgs branch ${\cal M}_{n,1}$ is 
precisely
the symmetric product $Y_n=S^n{\bf R}^4$  
of $n$ copies of ${\bf R}^4$, with the obvious flat
metric.  At first sight this strongly suggests that the conformal field
theory of the Higgs branch would be simply the conventional orbifold conformal
field theory 
with target space $Y$.  This assumption, however, leads to
 contradictions.  
If the conformal field theory of the Higgs branch
is simply an orbifold
obtained by dividing the symmetric product of $n$ copies of {\it anything} by
the symmetric group $S_n$, then the spectrum of the matrix string theory would
be -- by the standard logic of matrix string theory \refs{\motl - \dvv}
-- a free Fock space of one particle states.  Moreover, in this case the one
particle states in question would be,  essentially as guessed in \ref\olddvv{
R. Dijkgraaf, E. Verlinde, and H. Verlinde, ``BPS Quantization Of The 
Fivebrane,'' Nucl. Phys. {\bf B486} (1997) 89, hep-th/9604055.
}, the states of the six-dimensional light-cone Green-Schwarz
superstring.  But the Green-Schwarz superstring in six-dimensions is not 
Lorentz-invariant, six not being the critical dimension.  So whatever
the theory of a single fivebrane (at vanishing Type IIA string coupling 
constant) 
may be, it cannot have the spectrum of the
light-cone Green-Schwarz superstring.  

A resolution of this problem was proposed in \newdvv.  
The proposal (as adapted to the language of the present discussion) is that
the Higgs branch of the $U(n)$ theory with $k=1$ is not 
precisely the orbifold conformal field theory
 but is a sigma model whose target space is a
blow-up of $Y$.  The idea, to be more precise,
is that the locus that is blown up is the locus in $Y$ in which {\it two}
of the $n$ points in ${\bf R}^4$ coincide.  This locus is of codimension four,
and its blowup is represented by marginal deformations of the conformal field
theory.  Blowup of ``higher order diagonals'' on which more than two points 
coincide
would by contrast be represented by irrelevant operators.  Since we are blowing
up a locus where just two points coincide, the singularity that is being blown
up is a ${\bf Z}_2$ or $A_1$ orbifold singularity; the local behavior of the
conformal field theory near such a singularity
can in fact be described by a $U(1)$ theory with two
charged hypermultiplets (of equal charge), as in \witten.  Geometrically,
the blowup creates a two-cycle $C$ (which is topologically a two-sphere).

The proposal to describe the fivebrane via a sigma model whose target space
is this blowup
cannot be precisely correct.  The blowup of
$Y$ would violate the rotation symmetry of ${\bf R}^4$ (breaking $SO(4)$ to 
$SU(2)
\times U(1)$); but this $SO(4)$, which rotates the four transverse dimensions
of the fivebrane, is an exact symmetry of the fivebrane theory.  Even
more specifically, we saw in the last
subsection that the Higgs branch of the $U(n)$ theory with $k=1$ is equal on 
the nose
to 
the symmetric product $Y$, and is not a blowup of $Y$.

However, a variant of the proposal in \newdvv\ does seem to be viable.  The 
three blowup modes of the orbifold conformal field theory
associated with blowup of the codimension four singular
locus in $Y$ have a supersymmetric completion which is a fourth marginal 
operator
(like the blowup modes this operator is a twist field of the orbifold theory).
The coupling constant
associated with this  operator would be interpreted after blowup as a 
world-sheet theta angle; before the blowup, it is hard to interpret this
coupling as a worldsheet theta  angle (since the two-cycle $C$ 
is collapsed at the orbifold fixed point), but it  can be interpreted as in
\witten\ as the theta angle of a $U(1)$ gauge theory that is used in an 
effective
description near the $A_1$ orbifold singularity. 

A deformation of the theta angle does not violate the $SO(4)$ transverse 
rotation
symmetry of the fivebrane, but a generic such deformation does violate the
left-right symmetry on the worldsheet of the $1+1$-dimensional theory.  
There are two values of $\theta$, namely $\theta=0$ and $\theta=\pi$, which
do respect this left-right symmetry and in fact respect all symmetries 
that should be manifest for the light cone fivebrane.
An important fact in this subject is that the conventional orbifold theory
-- the one that can be expressed in terms of free fields -- is the theory
at $\theta=\pi$ \ref\aspinwall{P. Aspinwall, ``Enhanced Gauge Symmetry And
K3 Surfaces,''  Phys. Lett. {\bf B357} (1995) 329, hep-th/9507012.}.
The fivebrane theory cannot be described by the conventional orbifold, for
reasons already explained, so $\theta=\pi$ is ruled out.  We are thus left
with one option along these lines: 
the conformal field theory on the $k=1$ Higgs branch is
not the usual orbifold, but differs from it by having $\theta=0$ instead of
$\theta=\pi$.  It seems that this is the appropriate version of the conjecture
made in \newdvv.

As encouraging hints supporting this interpretation, we may note
the following:

(1) There is a unique and fairly elegant candidate, an improvement over
the state of affairs in \newdvv\ where the blowup parameters were not 
specified.

(2) As seen in \witten, $\theta=0$ is the one case where the deformed orbifold
is described by a theory that also has a Coulomb branch.  We are certainly
dealing in the fivebrane problem with a theory that has a Coulomb branch,
so this fact fits nicely.

(3) Finally, though it can be seen in an effective $U(1)$ theory near the
singularity, the theta angle in question is not naturally embedded in the 
underlying
$U(n)$ gauge theory.  It is thus very natural that the underlying $U(n)$ theory
would automatically lead to $\theta=0$ and not $\theta=\pi$ as the theory
on the Higgs branch.

Even if it is true that the fivebrane is described by the orbifold theory 
deformed
to $\theta=0$, it is not at all clear how much computational
power concerning fivebranes this will yield.  The $\theta=0$ theory is probably
strongly coupled and
 difficult to understand.  Since this description does differ from the
soluble orbifold by a marginal deformation (albeit one that violates the
six-dimensional Lorentz invariance) it is plausible that
it could be used as the basis for a counting of BPS states roughly along
lines guessed in \olddvv.  However, all this discussion depended upon assuming
that the $n=k=1$ theory has a quantum Higgs vacuum.

\bigskip\noindent{\it Extension To $k>1$}

\nref\duff{M. Duff,
``Strong/Weak Coupling Duality From The Dual String,'' Nucl. Phys. {\bf B442}
(1995) 47, hep-th/9501030.}
\nref\dynamics{E. Witten, ``String Theory Dynamics In Various Dimensions,''
Nucl. Phys. {\bf B443} (1995) 85.}
\nref\verlinde{E. Verlinde, 
``Global Aspects Of Electric-Magnetic Duality, Nucl. Phys. {\bf B455}
(1995) 211, hep-th/9506011.}
\nref\vafa{A. Klemm, W. Lerche, P. Mayr, C. Vafa, and N. Warner,
``Selfdual Strings And $N=2$ Supersymmetric Field Theory,''
Nucl. Phys. {\bf B477} (1996) 746, hep-th/9604034.}
\nref\newwitten{E. Witten, ``Branes And The Dynamics Of QCD,'' hep-th/9706109.}
We will now discuss the interactions in the case of
$k$ parallel fivebranes with $k>1$.  
In this case, less precise statements can be made, but at least we do not
need any doubtful assumptions about existence of a quantum Higgs branch.

To analyze this case, we must understand the Higgs branch
of the $U(n)$ theory with $k$ hypermultiplets in the fundamental 
representation.
In other words, we must understand the conformal field theory of the Higgs
branch ${\cal M}_{n,k}$.  A partial answer is provided by the result
explained in section 3.2 that ${\cal M}_{n,k}$ is birational to a symmetric
product of $n$ copies of ${\cal M}_{1,k}$.  However, this result  
was not accompanied by a useful description of the birational
transformation involved.  (Somewhat more detail can be found in \hurtu.) 
Ideally, one would like to understand the $U(n)$
theory with $k$ extra hypermultiplets as a marginal deformation of an orbifold
of $n$ copies of a $U(1)$ theory with $k$ charged hypermultiplets
(and a free hypermultiplet, corresponding to the adjoint representation of 
$U(1)$).
This would be a more far-reaching version of the relationship between the
moduli spaces ${\cal M}_{n,k}$ and ${\cal M}_{1,k}$.  It is not at all
clear, however, that such a description should be expected.

What could one learn from such a description, if it does exist?
There are many indications 
\refs{\duff-\verlinde,\witten,\vafa,\newwitten} that duality and dynamics of
four-dimensional gauge theories can be profitably derived from six dimensions.
In particular, a large $k$ expansion of the fivebrane 
theory might well lead in time to a $1/k$ expansion of $SU(k)$ gauge theory
in four dimensions, perhaps ultimately shedding great light on QCD.
The relation of  the moduli space ${\cal M}_{n,k}$
to a symmetric product of $n$ copies of ${\cal M}_{1,k}$ is a hint that such
an expansion may exist; for
${\cal M}_{1,k}$ is the Higgs branch of a theory ($U(1)$ with $k$ charged
hypermultiplets) that does have a $1/k$ expansion, as was emphasized in 
section 2.

\def\T{{\bf T}}
\bigskip\noindent{\it Compactification On ${\bf T}^4$}

\nref\maciocia{A. Maciocia,  
``Generalized Fourier-Mukai Transform,'' J. reine angew. Math. {\bf 480}
(1996) 197, 
alg-geom/9705001, ``Fourier-Mukai Transforms For Abelian Varieties
And Moduli Of Stable Bundles,'' Edinburgh University preprint (1996).}
\nref\friedman{R. Friedman, ``Vector Bundles And $SO(3)$ Invariants
For Elliptic Surfaces,'' J. Amer. Math. Soc. {\bf 8} (1995) 29,
alg-geom/9307002, ``Vector Bundles And $SO(3)$ Invariants For
Elliptic Surfaces III: The Case Of Odd Fiber Degree,'' alg-geom/9308004.}
In \newdvv, the compactification of the four transverse dimensions of the 
fivebrane on a four-torus $\T^4$ was considered.  It was proposed 
that in this case, the matrix string theory of $k$ fivebranes is a sigma model
whose target space is the moduli space of instanton number $n$ solutions of
$U(k)$ Yang-Mills theory on $\T^4$.  It is actually true 
\refs{\maciocia,\friedman}
that, at least with some
restrictions on the flat metric on the $\T^4$ and the Chern classes of the 
instanton
bundle, this moduli space is a blowup
of an orbifold which is a
 symmetric product $S^{kn}\T^4$. The idea in \newdvv\ is that interactions
would be generated by the blowup (and perhaps also, in view of what has been 
said
above, by a shift in the theta angle of the orbifold).  
While this proposal is perhaps
plausible, it would be beyond the scope of the present paper to analyze it,
since the logic of our discussion limits us to fivebranes on a transverse
${\bf R}^4$.
In fact, the experience in the matrix model is that compactification brings
many surprises, though perhaps in the case of fivebranes these do not affect
the relationship between the light cone sigma model and the instanton moduli 
space.

Note that there does not appear to be a known birational
relation of the moduli space of
$n$ $U(k)$  instantons on ${\bf R}^4$ to a symmetric product of $nk$ copies
of ${\bf R}^4$.\foot{Though not obviously relevant to proposals about fivebrane
interactions, a birational relation of ${\cal M}_{n,k}$
to a symmetric product of $n$
copies of ${\bf R}^{4k}$ can be deduced using the description
in section 3.2.  This is a consequence of the fact that ${\cal M}_{n,k}$ is
birational to a symmetric product of $n$ copies of ${\cal M}_{1,k}$,
and ${\cal M}_{1,k}$ is
birational to ${\bf C}^{2k}$.  The latter statement
can be proved by using the ${\bf C}^*$ action to set one of the $A_\alpha$, 
say $A_1$, to 1, and then using the equation $\sum
A_\alpha B_\alpha=0$ to eliminate $B_1$, leaving $u,v$ and the other $A$'s and
$B$'s to generically parametrize ${\cal M}_{1,k}$.}  
Such a relation, at any rate, could apparently not commute
with the action on the instanton moduli space of the global symmetry group 
$SU(k)$.
This group acts because (as explained in the footnote in section 3.2) in 
defining
the instanton moduli space, we consider two instantons to be gauge-equivalent
only if they are equivalent by a gauge transformation that is the identity at
infinity.  After dividing by this equivalence, one has a global
action of the $U(k)$ gauge transformations at infinity on the instanton moduli 
space.
By contrast, the group $U(k)$ does not act in any evident  fashion
on the symmetric product
$S^{nk}{\bf R}^4$.  

There is no obvious and useful relation between instanton
moduli spaces on $\T^4$ and on ${\bf R}^4$.  They differ because on $\T^4$
one has Wilson lines that have no analog on ${\bf R}^4$, and 
because in defining instanton moduli space on ${\bf R}^4$, one divides only
by those gauge transformations that equal the identity at infinity, a condition
that has no analog on ${\bf T}^4$.  A linear sigma model whose Higgs branch
is a rough analog of instanton moduli space on ${\bf T}^4$ would
be a $U(n)\times U(k)$ gauge theory with hypermultiplets consisting
of a copy of the adjoint
representation of each of the two factors together with an $(n,k)$ 
hypermultiplet.

\bigskip

I would like to thank R. Dijkgraaf, M. Douglas, D. Freed, N. Seiberg, S. Sethi,
E. Verlinde, H. Verlinde, and S. Wadia
for discussions and to thank the University of Amsterdam and the organizers
of {\it Strings '97} for their hospitality.

\listrefs
\end